%% file: intermediate_arxiv_final.tex
\documentclass[journal]{IEEEtran}
\usepackage{amsmath,amsfonts, amssymb, mathrsfs}
\usepackage[dvips]{graphicx}
\newtheorem{theorem}{Theorem}
\newtheorem{lemma}{Lemma}

\def\EE{\mathbb E}

\def\SNR{\text{SNR}}
\def\et{\quad \text{and} \quad}
\def \PP{\mathbb P}
\def \Tr{\text{Tr}}

\begin{document}
\title{Information Theoretic Operating Regimes of Large Wireless Networks}
\author{Ayfer {\"O}zg{\"u}r, Ramesh Johari,~\IEEEmembership{Member,~IEEE}, David Tse,~\IEEEmembership{Fellow,~IEEE} and Olivier L\'ev\^eque,~\IEEEmembership{Member,~IEEE}
\thanks{
The work of Ayfer {\"O}zg{\"u}r was supported by Swiss NSF grant Nr 200020-118076. The work of Ramesh Johari was supported by the Defense Advanced Research Projects Agency under the 
ITMANET program, and the U.S. National Science Foundation under grant 0644114. The work of David Tse was supported by the U.S. National Science Foundation via an ITR grant: ``The 3R's of Spectrum Management: Reuse, Reduce and Recycle''.
The material in this paper was presented in part at the IEEE Symposium on Information Theory, Toronto, July 2008.}
\thanks{A.~\"Ozg\"ur and O.~L\'ev\^eque are with the Ecole Polytechnique F\'ed\'erale de Lausanne, Facult\'e Informatique et Communications, Building INR, Station 14, CH - 1015 Lausanne, Switzerland (e-mails: \{ayfer.ozgur,olivier.leveque\}@epfl.ch).}
\thanks{R.~Johari is with the Stanford University, Department of Management Science and Engineering, Stanford, CA 94305, USA (e-mail: ramesh.johari@stanford.edu).}
\thanks{D.~Tse is with the University of California at Berkeley, Department of EECS, Berkeley, CA 94720, USA, (e-mail: dtse@eecs.berkeley.edu).}}

\maketitle

\begin{abstract}
In analyzing the point-to-point wireless channel, insights about two qualitatively
different operating regimes---bandwidth- and power-limited---have proven
indispensable in the design of good communication schemes. In this paper,
we propose a new scaling law formulation for wireless networks that allows
us to develop a theory that is analogous to the point-to-point case. We identify
fundamental operating regimes of wireless networks and derive
architectural guidelines for the design of optimal schemes. 

Our analysis shows that in a given wireless network with arbitrary size, area, power, bandwidth,
etc., there are three parameters of importance: the
short-distance SNR, the long-distance SNR, and the power path loss
exponent of the environment.  Depending on these parameters we identify four qualitatively different regimes. One of these regimes is especially interesting since it is fundamentally a consequence of the heterogeneous nature of links in a network and does not occur in the point-to-point case; the network capacity is {\em both } power and bandwidth limited. This regime has thus far remained hidden due to the limitations of the existing
formulation. Existing schemes, either multihop transmission or hierarchical
cooperation, fail to achieve capacity in this regime; we propose a new hybrid scheme that achieves capacity.
\end{abstract}


\section{Introduction}

The classic capacity formula $C = W \log_2 (1+ P_r/N_0W)$ bits/s of
a point-to-point AWGN channel with bandwidth $W$ Hz, received power
$P_r$ Watts, and white noise with power spectral density $N_0/2$
Watts/Hz plays a central role in communication system design. The
formula not only quantifies exactly the performance limit of
communication in terms of system parameters, but perhaps more
importantly also identifies two fundamentally different operating
regimes. In the power-limited (or low SNR)  regime, where $\SNR:=
P_r/N_0 W \ll 0$ dB, the capacity is approximately linear in the
power and the performance depends critically on the power available
but not so much on the bandwidth. In the bandwidth-limited (or high
SNR) regime, where $\SNR \gg 0$ dB, the capacity is approximately
linear in the bandwidth and the performance depends critically on
the bandwidth but not so much on the power. The regime is determined
by the interplay between the amount of power and degrees of freedom
available. The design of good communication schemes is primarily
driven by the parameter regime one is in.

Can analogous operating regimes be identified for ad hoc wireless
networks,  with multiple source and destination pairs and nodes
relaying information for each other? To address this question, we
are confronted with several problems. First, we have no exact
formula for the capacity of networks, even in the simplest case of a
single source-destination pair plus one relay. Second, unlike in the
point-to-point case, there is no single received SNR parameter in a
network. The channels between nodes closer together can be in the
high SNR regime while those between nodes farther away can be in the
low SNR regime.

One approach to get around the first problem is through the scaling
law formulation. Pioneered by Gupta and Kumar \cite{GK00}, this
approach seeks not the exact capacity of the network but only how it
{\em scales} with the number of nodes in the network and the number
of source-destination pairs. The capacity scaling turns out to
depend critically on how the area of the network scales with the
number of nodes. Two network models have been considered in the
literature. In {\em dense} networks \cite{GK00,AS06,OLT07}, the area is fixed while the
density of the nodes increases linearly with the number of nodes. In
{\em extended} networks \cite{XK04,JVK04,LT05,XK06,AJV06,OLP06,OLT07}, the area grows linearly with the number of
nodes while the density is fixed. For a given path loss exponent,
the area of the network determines the amount of power that can be
transferred across the network and so these different scalings
couple the power transferred and the number of nodes in different
ways.

There are two significant limitations in using the existing scaling
law results to identify fundamental operating regimes of ad hoc
networks. First, the degrees of freedom available in a network
depend on the number of nodes in addition to the the amount of
bandwidth available. By {\em a priori} coupling the power
transferred in the network with the number of nodes in specific
ways, the existing formulations may be missing out on much of the
interesting parameter space. Second, neither dense nor extended
networks allow us to model the common scenario where the channels
between different node pairs can be in different SNR regimes. More
concretely, let us interpret a channel to be in high SNR in a large
network if the SNR goes to infinity with $n$, and in low SNR if the SNR
goes to zero with $n$.\footnote{We interpret a channel in both high and low SNR, if the SNR does
not depend on $n$.} Then it can be readily verified that in dense
networks, the channels between all node pairs are in the high SNR
regime, while in extended networks, the channels between all node
pairs are in the low SNR regime.

In this paper, we consider a generalization that allows us to
overcome these two limitations of the existing formulation. Instead
of considering a fixed area or a fixed density, we let the area of
the network scale like $n^\nu$ where $\nu$ can take on any real
value. Dense networks correspond to $\nu =0$ and extended networks
correspond to $\nu =1$. By analyzing the problem for all possible
values of $\nu$, we are now considering all possible interplay
between power and degrees of freedom. Note that in networks where
$\nu$ is strictly between $0$ and $1$, channels between nodes that
are far away will be at low SNR while nodes that are closer by will
be at high SNR. Indeed, the distance between nearest neighbors is of
the order of $\sqrt{A/n} = n^{(\nu-1)/2}$ and, assuming a path loss
exponent of $\alpha$, the received SNR of the transmitted signal
from the nearest neighbor scales like $n^{\alpha(1-\nu)/2}$, growing
with $n$. On the other hand, the received SNR of the transmitted
signal from the {\em farthest} nodes scales like $(\sqrt{A})^{-\alpha}
= n^{-\alpha \nu/2}$, going to zero with $n$. Note that scaling the
area by $n^{\nu}$ is completely equivalent to scaling the nearest
neighbor $\SNR$ as $n^\beta$, where $\beta := \alpha(1-\nu)/2$.
Since $\SNR$ is a physically more relevant parameter in designing
communication systems, we will formulate the problem as scaling
directly the nearest neighbor $\SNR$.

\bigbreak
The main result of this paper is as follows. Consider $2n$ nodes
randomly located in an area $2A$ such that the received SNR for a
transmission over the typical nearest neighbor distance of
$\sqrt{A/n}$ is $\SNR_s:= n^\beta$. The path loss exponent is $\alpha
\ge 2$. Each transmission goes through an independent uniform phase
rotation. There are $n$ source and destination pairs, randomly
chosen, each demanding the same rate. Let $C_n(\alpha,\beta)$ denote the total capacity of the network, which is the highest achievable sum rate, in bits/s/Hz and its scaling exponent be defined as,
\begin{equation}\label{def:scexp}
e(\alpha, \beta)  :=
\lim_{n \rightarrow \infty} \frac{\log C_n(\alpha,\beta)}{\log n}. 
\end{equation}

\bigbreak
The following theorem is the main result of this paper.
\begin{theorem}\label{thm1}
The scaling exponent $e(\alpha, \beta)$ of  the total capacity $C_n(\alpha,\beta)$ is given by
\begin{equation}
e(\alpha,\beta) = \left \{ \begin{array}{ll} 1 &  \beta \geq \alpha/2 -1 \\
2-\alpha/2 + \beta  & \beta < \alpha/2 - 1 \mbox{ and } 2 \le
\alpha \le 3 \\
1/2 + \beta & \beta \le 0 \mbox{ and } \alpha > 3\\
1/2 + \beta/(\alpha-2) &  0 < \beta < \alpha/2 - 1 \mbox{ and }
\alpha > 3.\label{eq:main}\end{array} \right .
\end{equation}
\end{theorem}
\bigbreak

Note that dense networks correspond to $\beta = \alpha/2$, with an
exponent $e(\alpha,\alpha/2) = 1$ (first case), and extended
networks correspond to $\beta = 0$, with an exponent equal to:
$$
e(\alpha,0) = \left
\{ \begin{array}{ll} 2 - \alpha/2 & 2 \le \alpha  \le 3 \\
1/2 & \alpha > 3
\end{array} \right .
$$
(second and third cases respectively). These special cases are the
main results of \cite{OLT07}. Observe that in the general case the
scaling exponent $e(\alpha,\beta)$ depends on the path loss exponent
$\alpha$ and the nearest neighbor $\SNR$ exponent $\beta$ {\em
separately}, so the general result cannot be obtained by a simple
re-scaling of distances in the dense or extended model.

To interpret the general result (\ref{eq:main}) and to compare it to
the point-to-point scenario, let us re-express the result in terms
of system quantities. Recall that $\SNR_s$ is the SNR over the smallest scale in the network, which is the typical nearest neighbor distance. Thus,
\begin{equation}\label{SNRs}
 \SNR_s = n^\beta = \frac{P_r}{N_0 W},
\end{equation}
where $P_r$ is the received power from a node at the typical nearest neighbor distance $\sqrt{A/n}$ and $W$ Hz is the channel bandwidth. Let us also define the SNR over the largest scale in the network, the diameter $\sqrt{A}$, to be
\begin{equation}\label{SNRl}
\SNR_l:=n\,\frac{n^{-\alpha/2}P_r}{N_0 W}=n^{1-\alpha/2+\beta}, 
\end{equation}
where $n^{-\alpha/2}P_r$ is the received power from a node at distance diameter of the network. The result (\ref{eq:main}) can be used to give the following approximation to
the total capacity $C$,  in  bits/s: \footnote{Note that $C=W\,C_n(\alpha,\beta)$.}

\begin{equation}
\label{eq:approx}
C \approx  \left \{ \hspace*{-0.2cm}\begin{array}{ll} n W &  \SNR_l \gg 0 \mbox { dB} \\
n^{2-\alpha/2} P_r/N_0 & \SNR_l \ll 0 \mbox { dB and } 2\le\alpha \le 3 \\
\sqrt{n} P_r/N_0  & \SNR_s \ll 0  \mbox{ dB and } \alpha > 3\\
\sqrt{n} W^{\frac{\alpha -3}{\alpha -2}} (P_r/N_0)^{\frac{1}{\alpha
-2}} & \SNR_l \ll 0 \mbox{ dB},\, \SNR_s \gg 0 \mbox{ dB}\\
& \mbox {and } \alpha > 3.
\end{array} \right .
\end{equation}
Note two immediate observations in (\ref{eq:approx}). First, there are two SNR parameters of interest in networks, the short and the long distance SNR's, as opposed to the point-to-point case where there is a single SNR parameter. Second, the most natural way to measure the long-distance SNR in networks is not the SNR of a pair separated by a distance equal to the diameter of the network, but it is $n$ times this quantity as defined in (\ref{SNRl}). Note that there are order $n$ nodes in total located at a diameter distance to any given node in the network, hence $n$ times the SNR between farthest nodes is the total SNR that can be transferred to this node across this large scale. On the other hand a node has only a constant number of nearest neighbors, and hence the short-distance SNR in (\ref{SNRs}) is simply the SNR between a nearest neighbor pair.

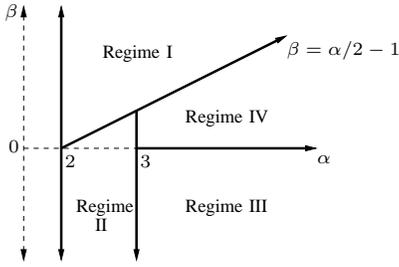
\begin{figure}[tbp]
\begin{center}
\input{fourregimes.pstex_t}
\end{center}
\caption{The four operating regimes. The optimal schemes in these regimes are I-Hierarchical Cooperation, II-Bursty Hierarchical Cooperation, III-Multihop, IV- Multihop MIMO Hierarchical Cooperation.}
 \label{fig:phase_diagram}
\end{figure}

The four regimes in (\ref{eq:approx}) are shown in Figure \ref{fig:phase_diagram}. 
In Regime-I, the performance is achieved by
hierarchical cooperation and long range MIMO transmission, the
scheme introduced in \cite{OLT07}. At the highest level of
hierarchy, clusters of size almost order $n$ communicate via MIMO,
at distance the diameter of the network. The quantity $\SNR_l$ corresponds to the total
received SNR at a node during these MIMO transmissions. Since this quantity is larger than $0$ dB,
the long range MIMO transmissions, and hence the performance of the
network, are in the bandwidth limited regime, with performance
roughly linear in the bandwidth $W$. The performance is linear in
the number of nodes, implying that interference limitation is
removed by cooperation, at least as far as scaling is concerned.
Performance in this regime is qualitatively the same as that in
dense networks.

In all the other regimes, the total long-range received SNR is less
than $0$ dB. Hence we are power-limited and the transfer of power
becomes important in determining performance. In Regime-II, i.e., when $\alpha \le 3$,
signal power decays slowly with distance, and the total power
transfer is maximized by long-range MIMO transmission. This
performance can be achieved by bursty hierarchical cooperation 
with long-range MIMO, much like in extended networks. 

When $\alpha > 3$, signal power decays fast with distance, and the
transfer of power is maximized by short-range communications. If the
nearest-neighbor $\SNR \ll 0$ dB (Regime-III), these transmissions are in the
power-efficient regime and this power gain translates linearly into
capacity, so nearest-neighbor multihop is optimal. This is indeed
the case in extended networks, and hence nearest-neighbor multihop
is optimal for extended networks when $\alpha > 3$.

The most interesting case is the fourth regime, when $\alpha > 3$
and $0 < \beta < \alpha/2 -1$. This is the case when $\SNR_s
\gg 0 $ dB, so nearest-neighbor transmissions are bandwidth-limited and not power-efficient
in translating the power gain into capacity gain. There is the potential
of increasing throughput by spatially multiplexing transmission via
cooperation within clusters of nodes and performing distributed
MIMO. Yet, the clusters cannot be as large as the size of the
network since power attenuates rapidly for $\alpha > 3$. 

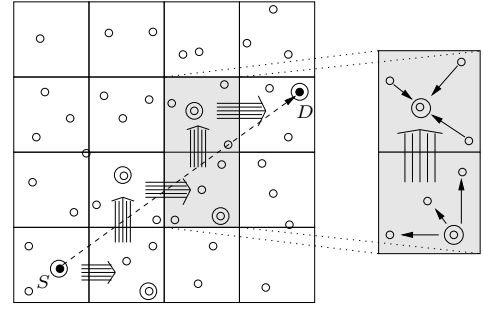
\begin{figure}[tbp]
\begin{center}
\input{multihopMIMO.pstex_t}
\end{center}
\caption{The figure illustrates the optimal scheme in Regime IV which is based on cooperating locally and multihopping globally. Note that packets are transmitted by multihopping on the network level and each hop is realized with distributed MIMO transmissions combined with hierarchical cooperation.}
\label{fig:mimo_multihop}
\end{figure}

Indeed, it turns out that the optimal scheme in this regime is to cooperate
hierarchically within clusters of an intermediate size, perform MIMO
transmission between adjacent clusters and then multihop across
several clusters to get to the final destination. (See Figure
\ref{fig:mimo_multihop}). The optimal cluster size is chosen such
that the received SNR in the MIMO transmission is at $0$ dB. Any
smaller cluster size results in power inefficiency. Any larger
cluster size reduces the amount of power transfer because of the
attenuation. Note that the two extremes of this architecture are
precisely the traditional multihop scheme, where the cluster size is
$1$ and the number of hops is $\sqrt{n}$, and the long-range
cooperative scheme, where the cluster size is of order $n$ and the
number of hops is $1$. Note also that because short-range links are
bandwidth-limited and long-range links are power-limited, the
network capacity is {\em both} bandwidth and power-limited. Thus the
capacity is sensitive to both the amount of bandwidth and the amount
of power available. This regime is fundamentally a consequence of
the heterogeneous nature of links in a network and does not occur in
point-to-point links, nor in dense or extended networks.

The organization of the paper is as follows. In the following section we present our model in more detail. Section~\ref{sec:ub} derives a tight upper bound on the scaling exponent in (\ref{def:scexp}). Section~\ref{sec:ach} introduces schemes that achieve the upper bound presented in the previous section. The two sections together prove our main result in Theorem~\ref{thm1}. Section~\ref{sec:conc} contains our conclusions.

\section{Model}
There are $2n$ nodes uniformly and independently distributed in a rectangle of area $2\sqrt{A}\times\sqrt{A}$. Half of the nodes are sources and the other half are destinations. The sources and destinations are randomly paired up one-to-one without any consideration on node locations. Each source has the same traffic rate $R$ in bits/s/Hz to send to its destination node and a common average transmit power budget of $P$ Watts. The total throughput of the system is $T = n R$.

We assume that communication takes place over a flat channel of
bandwidth $W$ Hz around a carrier frequency of $f_c$, $f_c \gg W$.
The complex baseband-equivalent channel gain between  node $i$ and
node $k$ at time $m$ is given by:
\begin{equation}
\label{eq:ch_model}
H_{ik}[m] = \sqrt{G}\,r_{ik}^{-\alpha/2} \exp(j \theta_{ik}[m])
\end{equation}
where $r_{ik}$ is the distance between the nodes, $\theta_{ik}[m]$ is
the random phase at time $m$, uniformly distributed in $[0,2\pi]$ and
$\{\theta_{ik}[m]; 1\leq i\leq 2n, 1\leq k\leq 2n\}$ is a collection of independent identically distributed random processes. The
$\theta_{ik}[m]$'s and the $r_{ik}$'s are also assumed to be
independent. The parameters $G$ and $\alpha \ge 2$  are assumed to be constants;
$\alpha$ is called the power path loss exponent.

The path-loss model is based on the standard far-field assumption: we
assume that the distance $r_{ik}$ is much larger than the carrier wavelength
$\lambda_c$. When the distances are comparable or shorter than the carrier
wavelength, the simple path-loss model obviously does not hold anymore as
path loss can potentially become path ``gain''. Moreover, the phases
$\theta_{ik}[m]$ depend on the distance between the nodes modulo the carrier
wavelength and they can only be modeled as completely random and
independent of the actual positions of the nodes if the nodes' separation
is large enough. Indeed, a recent result \cite{FMM08} showed that, without making an  a
priori assumption of i.i.d. phases, the degrees of freedom are limited by the diameter
of the network (normalized by the carrier wavelength $\lambda_c$). This is a {\em
spatial} limitation and holds regardless of how many communicating nodes there are
in the network. This result suggests that as long as the number of nodes $n$ is
smaller than this normalized diameter, the spatial degrees of freedom limitation does
not kick in, the degrees of freedom are still limited by the number of nodes, and the
i.i.d. phase model is still reasonable.  For example, in a network with diameter $1$
km and carrier frequency $3$ GHz, the number of nodes should be of the order of $10^4$ or less
for the i.i.d. phase model to be valid.  While the present paper deals exclusively
with the standard i.i.d. random phase model, it would be interesting to apply our
new scaling law formulation to incorporate regimes where there is a spatial degrees of freedom
limitation as well.

Note that the channel is random, depending on the location of the
users and the phases. The locations are assumed to be fixed over the
duration of the communication. The phases are assumed to vary in a stationary ergodic manner (fast fading). We assume that the phases $\{\theta_{ik}[m]\}$ are known in a casual manner at all the nodes in the network. The signal received by node $i$ at time $m$ is given by
$$
Y_i[m]=\sum_{k\neq i}H_{ik}[m]X_k[m]+Z_i[m]
$$
where $X_k[m]$ is the signal sent by node $k$ at time $m$ and
$Z_i[m]$ is white circularly symmetric Gaussian noise of variance
$N_0$ per symbol.

\section{Cutset Upper Bound}\label{sec:ub}

\begin{figure}[tbp]
\begin{center}
\input{cutset_small.pstex_t}
\end{center}
\caption{The cut-set considered in Section~\ref{sec:ub}. The communication requests that pass across the cut from left to right are depicted in bold lines.}
\label{fig:cutset}
\end{figure}
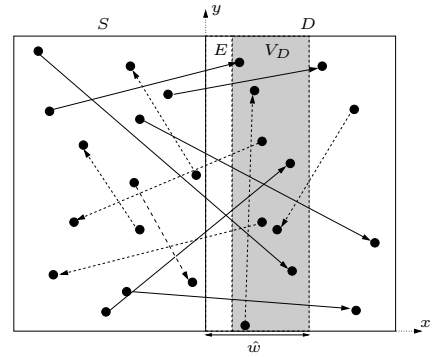

We consider a cut dividing the network area into two equal halves. We are interested in upper bounding the sum of the rates of communications $T_{L\rightarrow R}$ passing through the cut from left to right. These communications with source nodes located on the left and destination nodes located on the right half domain are depicted in bold lines in Fig.~\ref{fig:cutset}. Since the S-D pairs in the network are formed uniformly at random, $T_{L\rightarrow R}$ is equal to $1/4$'th of the total throughput $T$ w.h.p.\footnote{with high probability, i.e., with probability going to $1$ as $n$ grows.} The maximum achievable $T_{L\rightarrow R}$ in bits/s/Hz is bounded above by the capacity of the MIMO channel between all nodes $S$ located to the left of the cut and all nodes $D$ located to the right. Under the fast fading assumption, we have

\begin{equation}
T_{L\rightarrow R}
\leq \hspace*{-0.3cm}\max_{\substack{Q(H) \geq 0 \\ \EE(Q_{kk}(H)) \leq P,
\, \forall k\in S}} \hspace*{-0.5cm}\EE \left(\log \det(I +\frac{1}{N_0W} H Q(H) H^*) \right) \label{cutset}
\end{equation}
where
$$
H_{ik} = \frac{\sqrt{G} \; e^{j \, \theta_{ik}}}{r_{ik}^{\alpha/2}},
\quad k\in S, i\in D.
$$
The mapping $Q(\cdot)$ is from the set of possible channel realizations $H$ to the set of positive semi-definite transmit covariance matrices. The diagonal element $Q_{kk}(H)$ corresponds to the power allocated to the $k$th node for channel state $H$. Let us simplify notation by introducing
\begin{equation}\label{SNR_0}
\SNR_s:=\frac{GP}{N_0W(A/n)^{\alpha/2}}
\end{equation}
which can be interpreted as the average SNR between nearest neighbor nodes since $\sqrt{A/n}$ is the typical nearest neighbor distance in the network. Let us also rescale the distances in the network by this nearest neighbor distance, defining
\begin{equation}
\hat{r}_{ik}:=\frac{1}{\sqrt{A/n}}\, r_{ik} \et \hat{H}_{ik} := \frac{ e^{j \, \theta_{ik}}}{\hat{r}_{ik}^{\alpha/2}}. \label{dis_ext}
\end{equation}
Note that the first transformation rescales space and maps our original network of area $2\sqrt{A}\times \sqrt{A}$ to a network of area $2\sqrt{n}\times \sqrt{n}$, referred to as an extended network in the literature. Consequently, the matrix $\hat{H}$ defined in terms of the rescaled distances relates to such an extended network with area $2n$. We can rewrite (\ref{cutset}) in terms of these new variables as
\begin{equation}
T_{L\rightarrow R}
\leq \hspace*{-0.3cm}\max_{\substack{Q(\hat{H}) \geq 0 \\ \EE(Q_{kk}(\hat{H})) \leq 1,
\, \forall k\in S}} \hspace*{-0.5cm}\EE \left( \log \det(I + \SNR_{s}\,\hat{H} Q(\hat{H}) \hat{H}^*) \right). \label{cutset_simp}
\end{equation}
In order to upper bound (\ref{cutset_simp}), we will use an approach similar to the one developed in \cite[Sec.V-B]{OLT07} for analyzing the capacity scaling of extended networks. Note that although due the rescaling in (\ref{dis_ext}), $\hat{H}$ in (\ref{cutset_simp}) governs an extended network, the problem in (\ref{cutset_simp}) is not equivalent to the classical extended setup since here we do not necessarily assume $\SNR_{s}=1$. Indeed, we want to keep full generality and avoid such arbitrary assumptions on $\SNR_{s}$ in the current paper. Formally, we are interested in characterizing the whole regime $\SNR_{s}=n^\beta$ where $\beta$ can be any real number.

One way to upper bound (\ref{cutset_simp}) is through upper bounding the capacity by the total received SNR, formally using the relation 
\begin{equation}\label{up:tr}
\log \det(I + \SNR_{s}\,\hat{H} Q(\hat{H}) \hat{H}^*)\leq \Tr\left(\SNR_{s}\,\hat{H} Q(\hat{H}) \hat{H}^* \right). 
\end{equation}
The upper bound is tight only if the SNR received by each right-hand side node (each diagonal entry of the matrix $\SNR_{s}\,\hat{H} Q(\hat{H}) \hat{H}^*$) is small. (Note that the relation in (\ref{up:tr}) relies on the inequality $\log(1+x)\leq x$ which is only tight if $x$ is small.) In the extended setup, where $\SNR_{s}=1$, the network is highly power-limited and the received SNR is small, that is decays to zero with increasing $n$, for every right-hand side node. Using (\ref{up:tr}) yields a tight upper bound in that case. However, in the general case $\SNR_{s}$ can be arbitrarily large which can result in high received SNR for certain right-hand side nodes that are located close to the cut or even for all nodes, depending on how large exactly $\SNR_{s}$ is. Hence, before using (\ref{up:tr}) we need to distinguish between those right-hand side nodes that receive high SNR and those that have poor power connections to the left-hand side.

For the sake of simplicity in presentation, we assume in this section that there is a rectangular region located immediately to the right of the cut that is cleared of nodes. Formally, we assume that the set of nodes $E=\{i\in D: 0\leq \hat{x}_i\leq 1\}$ is empty, where $\hat{x}_i$ denotes the horizontal coordinate of the rescaled position $\hat{r}_i=(\hat{x}_i,\hat{y}_i)$ of node $i$. In fact, w.h.p this property does not hold in a random realization of the network. However, making this assumption allows us to exhibit the central ideas of the discussion in a simpler manner. The extension of the analysis to the general case (without this particular assumption) is given in Appendix~\ref{sec:app2}. 

Let $V_D$ denote the set of nodes located on a rectangular strip immediately to the right of the empty region $E$. Formally, $V_D=\{i\in D: 1\leq \hat{x}_i\leq \hat{w}\}$ where  $1\leq \hat{w}\leq \sqrt{n}$ and $\hat{w}-1$ is the rescaled width of the rectangular strip $V_D$. See Fig.~\ref{fig:cutset}. We would like to tune $\hat{w}$ so that $V_D$ contains the right-hand side nodes with high received SNR; i.e., those with received SNR larger than a threshold, say $1$. Note however that we do not yet know the covariance matrix $Q$ of the transmissions from the left-hand side nodes, which is to be determined from the maximization problem in (\ref{cutset_simp}). Thus, we cannot compute the received SNR of a right-hand side node. For the purpose of determining $V_D$ however, let us arbitrarily look at the case when $Q$ is the identity matrix and define the received SNR of a right-hand side node $i\in D$ when left-hand side nodes are transmitting {\em independent} signals at full power to be
\begin{equation}\label{SNRi}
\SNR_i := \frac{P}{N_0W}\,\sum_{k\in S} |H_{ik}|^2\,=\,\SNR_{s}\,\sum_{k\in S} |\hat{H}_{ik}|^2=\,\SNR_{s}\,\hat{d}_i.
\end{equation}
where we have defined 
\begin{equation}\label{di}
\hat{d}_i:=\sum_{k\in S} |\hat{H}_{ik}|^2. 
\end{equation}
Later, we will see that this arbitrary choice of identity covariance matrix is indeed a reasonable one. A good approximation for $\hat{d}_i$ is
\begin{equation}
\label{di_app}
\hat{d}_i\approx \hat{x}_i^{2-\alpha}
\end{equation}
where $\hat{x}_i$ denotes the rescaled horizontal coordinate of node $i$. (See \cite[Lemma~5.4]{OLT07}.) Using (\ref{SNRi}) and (\ref{di_app}), we can identify three different regimes and specify $\hat{w}$ accordingly:
\begin{itemize}
\item[1)] If $\SNR_{s}\,\geq\,n^{\alpha/2-1}$, then $\SNR_i\gtrsim 1,\,\forall i\in D$. Thus, let us choose $\hat{w}=\sqrt{n}$ or equivalently $V_D=D$.
\medbreak
\item[2)] If $\SNR_{s}\,<\,1$, then $\SNR_i\lesssim 1,\,\forall i\in D$. Thus, let us choose $\hat{w}=1$ or equivalently $V_D=\emptyset$.\footnote{Note that this is when we use the earlier assumption of an empty strip $E$ of width $1$. Without the assumption, we would need to choose $\hat{w}<1$ in this part.}
\medbreak
\item[3)] If $1\leq\SNR_{s}\,<\,n^{\alpha/2-1}$, then let us choose
$$
\hat{w}=\left \{ \begin{array}{ll} \sqrt{n} & \text{if}\qquad \alpha =2 \\
\SNR_{s}^\frac{1}{\alpha-2} & \text{if}\qquad\alpha > 2
\end{array} \right .
$$
so that  we ensure $\SNR_i \gtrsim 1, \,\forall i\in V_D$.
\end{itemize}
\medbreak
We now would like to break the information transfer from the left-half domain $S$ to the right-half domain $D$ in (\ref{cutset_simp}) into two terms. The first term governs the information transfer from $S$ to $V_D$. The second term governs the information transfer from $S$ to the remaining nodes on the right-half domain, i.e., $D\setminus V_D$. Recall that the characteristic of the nodes $V_D$ is that they have good power connections to the left-hand side, that is the information transfer from $S$ to $V_D$ is not limited in terms of power but can be limited in degrees of freedom. Thus, it is reasonable to bound the rate of this first information transfer by the cardinality of the set $V_D$ rather than the total received SNR. On the other hand, the remaining nodes $D\setminus V_D$ have poor power connections to the left-half domain and the information transfer to these nodes is limited in power, hence using (\ref{up:tr}) is tight. Formally, we proceed by applying the generalized Hadamard's inequality which yields
\begin{align*}
\log \det(I &+ \SNR_s\hat{H} Q(\hat{H}) \hat{H}^*)\\
& \leq \log\det(I + \SNR_s\hat{H}_1 Q(\hat{H}) \hat{H}_1^*)\\
&\hspace*{0.5cm}+\log\det(I + \SNR_s\hat{H}_2 Q(\hat{H}) \hat{H}_2^*)
\end{align*}
where $\hat{H}_1$ and $\hat{H}_2$ are obtained by partitioning the original matrix $\hat{H}$: $\hat{H}_1$ is the rectangular matrix with entries $\hat{H}_{ik}, k\in S, i\in V_D$ and $\hat{H}_2$ is the rectangular matrix with entries $\hat{H}_{ik}, k\in S, i\in D\setminus V_D$. In turn, (\ref{cutset_simp}) is bounded above by
\begin{eqnarray}
\lefteqn{\hspace{-0.4cm}T_{L\rightarrow R}
\leq \hspace{-0.7cm}\max_{\substack{Q(\hat{H}_1) \geq 0 \\ \EE(Q_{kk}(\hat{H}_1)) \leq 1,
\, \forall k\in S}} \hspace{-0.7cm}\EE \left( \log \det(I + \SNR_s\hat{H}_1 Q(\hat{H}_1) \hat{H}_1^*) \right) \nonumber}\\
& + & \hspace{-1.2cm} \max_{\substack{Q(\hat{H}_2) \geq 0 \\ \EE(Q_{kk}(\hat{H}_2)) \leq 1,
\, \forall k\in S}} \hspace{-0.7cm}\EE \left( \log \det(I + \SNR_s\hat{H}_2 Q(\hat{H}_2) \hat{H}_2^*) \right)\label{cutset3}
\end{eqnarray}
The first term in (\ref{cutset3}) can be  bounded by considering the sum of the capacities of the individual MISO channels between nodes in $S$ and each node in $V_D$, 
\begin{eqnarray}
\lefteqn{\hspace{-0.5cm}\max_{\substack{Q(\hat{H}_1) \geq 0 \\ \EE(Q_{kk}(\hat{H}_1)) \leq 1,
\, \forall k\in S}} \hspace{-0.5cm}\EE \left( \log \det(I + \SNR_s\hat{H}_1 Q(\hat{H}_1) \hat{H}_1^*) \right)}\nonumber\\
& \leq  & \hspace{-0.3cm}\sum_{i\in V_D}\log(1+n\,\SNR_{s}\sum_{k\in S} |\hat{H}_{ik}|^2)\nonumber\\
& \leq  & \hspace{-0.3cm}(\hat{w}-1)\,\sqrt{n}\log n\,\log(1+n^{1+\alpha(1/2+\delta)}\,\SNR_{s})\label{ub1}
\end{eqnarray}
w.h.p. for any $\delta>0$, where we use the fact that for any covariance matrix $Q$ of the transmissions from the left-hand side, the SNR received by each node $i\in V_D$ is smaller than $n\,\SNR_{s} \hat{d}_i$ and $\hat{d}_i\leq n^{\alpha(1/2+\delta)}$ since the rescaled minimal separation between any two nodes in the network is larger than $\frac{1}{n^{1/2+\delta}}$ w.h.p. for any $\delta>0$. The number of nodes in $V_D$ is upper bounded by $(\hat{w}-1)\,\sqrt{n}\log n$ w.h.p.

The second term in (\ref{cutset3}) is the capacity of the MIMO channel between nodes in $S$ and nodes in $D\setminus V_D$. Using (\ref{up:tr}), we get 
\begin{align}
\max_{\substack{Q(\hat{H}_2) \geq 0 \\ \EE(Q_{kk}(\hat{H}_2)) \leq 1,\, \forall k\in S}} &\hspace*{-0.5cm}\EE \left( \log \det(I + \SNR_s\hat{H}_2 Q(\hat{H}_2) \hat{H}_2^*) \right)\nonumber\\
&\le \hspace*{-0.7cm}\max_{\substack{Q(\hat{H}_2) \geq 0 \\ \EE(Q_{kk}(\hat{H}_2)) \leq 1,\, \forall k\in S}} \hspace*{-0.7cm}\EE \left(\Tr\left(\SNR_s\hat{H}_2 Q(\hat{H}_2) \hat{H}_2^*\right)\right)\nonumber\\
&\le n^\epsilon\,\SNR_{tot}\label{ub2}
\end{align}
for any $\epsilon>0$ w.h.p, where
\begin{equation}\label{eq:snrtot}
\SNR_{tot}=\sum_{i\in D\setminus V_D} \SNR_i= \SNR_s \sum_{i\in D\setminus V_D}\hat{d}_i.
\end{equation}
Inequality (\ref{ub2}) is proved in \cite[Lemma 5.2]{OLT07} and is precisely showing that an identity covariance matrix is good enough for maximizing the power transfer from the left-hand side. Recall that $\SNR_i$ in (\ref{eq:snrtot}) has already been defined in (\ref{SNRi}) to be the received SNR of node $i$ under {\em independent} signalling from the left-hand side. Note that (\ref{eq:snrtot}) is equal to zero when $D\setminus V_D=\emptyset$ or equivalently when $\hat{w}=\sqrt{n}$. If $D\setminus V_D\neq\emptyset$, the last summation in (\ref{eq:snrtot}) can be approximated with an integral since nodes are uniformly distributed on the network area. Using also (\ref{di_app}), it is easy to derive the following approximation for the summation 
$$
\sum_{i\in D\setminus V_D}\hat{d}_i\,\approx\,\int_0^{\sqrt{n}}\int_{\hat{w}}^{\sqrt{n}} \hat{x}^{(2-\alpha)} d\hat{x}\,d\hat{y}.
$$ 
Here we state a precise result that can be found by straight forward modifications of the analysis in \cite{OLT07}. If $\hat{w}\neq \sqrt{n}$, we have
\begin{equation}
\SNR_{tot}\leq \left \{
\begin{array}{ll}
K_1\,\SNR_{s}\, n\, (\log n)^3 & \alpha =2 \vspace*{0.1cm}\\
K_1\,\SNR_{s}\, n^{2-\alpha/2}(\log n)^2  &2 < \alpha <3 \vspace*{0.1cm}\\
K_1\,\SNR_{s}\,\sqrt{n} \, (\log n)^3 & \alpha=3\vspace*{0.1cm} \\
K_1\,\SNR_{s}\,\hat{w}^{3-\alpha}\,\sqrt{n} \,(\log n)^2 & \alpha > 3. \end{array} \right.\label{SNRtot}
\end{equation}
where $K_1>0$ is a constant independent of $SNR_s$ and $n$.

Combining the upper bounds (\ref{ub1}) and (\ref{ub2}) together with our choices for $\hat{w}$ specified earlier, one can get an upper bound on $T_{L\rightarrow R}$ in terms of $\SNR_s$ and $n$. Here, we state the final result in terms of scaling exponents: Let us define
$$
\beta:=\lim_{n\to\infty}\frac{\log \SNR_s}{\log n}
$$
and
\begin{equation}
e(\alpha,\beta):=\lim_{n\to\infty}\frac{\log T}{\log n}\,=\,\lim_{n\to\infty}\frac{\log T_{L\rightarrow R}}{\log n}.\label{expo}
\end{equation}
We have,
\begin{equation}
e(\alpha, \beta) \leq \left \{ \begin{array}{ll} 1  & \beta \ge \alpha/2-1  \\
2-\alpha/2+\beta &  \beta < \alpha/2-1\,\, \text{and}\,\, 2\leq \alpha <3\\
1/2+\beta  & \beta \leq 0 \,\,\text{and}\,\, \alpha \ge 3\\
1/2+\beta/(\alpha-2) & 0 < \beta < \alpha/2-1 \,\,\text{and}\,\, \alpha \ge 3
\end{array} \right.\label{upp_expo}
\end{equation}
where we identify four different operating regimes depending on $\alpha$ and $\beta$. 

Note that in the first regime the upper bound (\ref{ub1}) is active with $\hat{w}=\sqrt{n}$ (or equivalently $V_D=D$) while (\ref{ub2}) is zero. The capacity of the network is limited by the degrees of freedom in an $n\times n$ MIMO transmission between the left and the right hand side nodes. In the second regime, (\ref{ub2}), with the corresponding upper bound being the second line in (\ref{SNRtot}), yields a larger contribution than (\ref{ub1}). The capacity is limited by the total received SNR in a MIMO transmission between the left-hand side nodes and $D\setminus V_D$. Note that this total received SNR is equal (in order) to the power transferred in a MIMO transmission between two groups of $n$ nodes separated by a distance of the order of the diameter of the network, i.e., $n^2\times (\sqrt{n})^{-\alpha}\times \SNR_s$. 

In the third regime, (\ref{ub2}) is active with $\hat{w}=1$ (or equivalently $V_D=\emptyset$) while (\ref{ub1}) is zero. The corresponding upper bound is the fourth line in (\ref{SNRtot}). Note that this is where we make use of the assumption that there are no nodes located at rescaled distance smaller than $1$ to the cut. Due to this assumption, the choice $\hat{w}=1$ vanishes the upper bound (\ref{ub1}) and simultaneously yields $K_1 \SNR_s \sqrt{n} (\log n)^2$ in the last line in (\ref{SNRtot}). If there were nodes closer than rescaled distance $1$ to the cut, we would need to choose $\hat{w}<1$ to vanish the contribution from (\ref{ub1}) which would yield a larger value for the term $K_1 \SNR_s \hat{w}^{3-\alpha}\sqrt{n} (\log n)^2$. The capacity in the third regime is still limited by the total SNR received by nodes in $D\setminus V_D$ ($=D$ now) but in this case the total is dominated by the SNR transferred between the nearest nodes to the cut, i.e., $\sqrt{n}$ pairs separated by the nearest neighbor distance, yielding $\sqrt{n}\times\SNR_s$.

The most interesting regime is the fourth one. Both (\ref{ub1}) and (\ref{ub2}) with the choice $\hat{w}=\SNR_s^{\frac{1}{\alpha-2}}$ yield the same contribution. Note that (\ref{ub1}) upper bounds the information transfer to $V_D$, the set of nodes that have bandwidth-limited connections to the left-hand side. This information transfer is limited in degrees of freedom. On the other hand, (\ref{ub2}) upper bounds the information transfer to $D\setminus V_D$, the set of nodes that have power-limited connections to the left-hand side. This second information transfer is power-limited. Eventually in this regime, the network capacity is both limited in degrees of freedom and power, since increasing the bandwidth increases the first term (\ref{ub1}) and increasing the power increases the second term (\ref{ub2}).

\section{Order Optimal Communication Schemes}\label{sec:ach}

In this section, we search for communication schemes whose performance meets the upper bound derived in the previous section. The derivation of the upper bound already provides hints on what these schemes can be: In the first two regimes, the capacity of the network is limited by the degrees of freedom and received SNR respectively, in a network wide MIMO transmission. The recently proposed hierarchical cooperation scheme in \cite{OLT07} is based on such MIMO transmissions so it is a natural candidate for optimality in these regimes. 

In the third regime, the information transfer between the two halves of the network is limited by the power transferred between the closest nodes to the cut. This observation suggests the following idea: if the objective is to transfer information from the left-half network to the right-half, then it is enough to employ only those pairs that are located closest to the cut and separated by the nearest neighbor distance. (The rest of the nodes in the network can undertake simultaneous transmissions suggesting the idea of spatial reuse.) In other words, the upper bound derivation suggests that efficient transmissions in this regime are the point-to-point transmissions between nearest neighbors. Indeed, this is how the well-known multihop scheme transfers power across the network so the multihop scheme arises as a natural candidate for optimality in the third regime. 

In the derivation of the upper bound for the fourth regime, we have seen that the two terms (\ref{ub1}) and (\ref{ub2}), governing the information transfer to $V_D$ and $D\setminus V_D$ respectively, yield the same contribution with the particular choice $\hat{w}=\SNR_s^{\frac{1}{\alpha-2}}$. Since the contributions of the two terms are equal (and since we are interested in order here) the derivation of the upper bound suggests the following idea: information can be transferred optimally from the left-half network to the right-half by performing MIMO transmission only between those nodes on both sides of the cut that are located up to $\hat{w}=\SNR_s^{\frac{1}{\alpha-2}}$ rescaled distance to the cut. Note that (\ref{ub1}) corresponds to the degrees of freedom in such a MIMO transmission. As in the case of multihop, we can have spatial reuse and allow the rest of the nodes in the network to perform simultaneous transmissions. Thus, the derivation of the upper bound  suggests that efficient transmissions in the fourth regime are MIMO transmissions at the scale $\hat{w}=\SNR_s^{\frac{1}{\alpha-2}}$. Combined with the idea of spatial reuse this understanding suggests to transfer information in the network by performing MIMO transmissions at the particular (local) scale of $\hat{w}=\SNR_s^{\frac{1}{\alpha-2}}$ and then multihopping at the global scale. This new scheme is introduced in Section~\ref{sec:ach:new}.\footnote{In a different context, a similar scheme has been suggested recently in an independent work \cite{NGS07}.}

\subsection{Known Schemes in the Literature}\label{sec:ach:old}

There are two fundamentally different communication schemes suggested for wireless networks in the literature: The multihop scheme and the hierarchical cooperation scheme. The multihop scheme is based on multihopping packets via nearest neighbor transmissions. Its aggregate throughput is well known to be
$$
T_{multihop}=\sqrt{n}\,\log\left(1+\frac{\SNR_s}{1+K_2\SNR_s}\right)
$$
w.h.p where $\log(1+\frac{\SNR_s}{1+K_2\SNR_s})$ is the throughput achieved in the nearest neighbor transmissions. $K_2\SNR_s$ is the interference from simultaneous transmissions in the network to noise ratio where $K_2>0$ is a constant independent of $n$ and $\SNR_s$. The factor $\sqrt{n}$ is the number of nearest neighbor transmissions that can be parallelized over a given cut. The scaling exponent $e_{multihop}(\alpha, \beta)$ of the multihop scheme (defined analogously to (\ref{expo})) is given by
\begin{equation}\label{expo_mh}
e_{multihop}(\alpha, \beta)=\left
\{ \begin{array}{ll} 1/2 & \beta > 0 \\
1/2+\beta & \beta \le 0
\end{array} \right .
\end{equation}
As can be expected, multihop only achieves the upper bound in (\ref{upp_expo}) in the third regime when $\beta\leq 0$ and $\alpha\geq 3$. In other words, when even the nearest neighbor transmissions in the network are power limited and signals attenuate sufficiently fast so that pairs located farther apart than the nearest neighbor distance cannot contribute to the power transfer effectively, the optimal strategy is to confine to nearest neighbor transmissions.
\medbreak
The second scheme for wireless networks in \cite{OLT07} is based on a hierarchical cooperation architecture that performs distributed MIMO transmissions between clusters of nodes. The overhead introduced by the cooperation scheme is small so that the throughput achieved by the distributed MIMO transmissions is not that different (at least in scaling sense) from the throughput of a classical MIMO system where transmit and receive antennas are collocated and can cooperate for free. Indeed, the aggregate throughput achieved by the scheme is almost equal to the rate of a MIMO transmission between two clusters of the size of the network $n$ and separated by a distance equal to the diameter of the network $\sqrt{A}$. More precisely,
\begin{equation}\label{hc}
T_{HC}\geq K_3\,n^{1-\epsilon}\,\log\left(1+ n\frac{GP}{N_0W\,(\sqrt{A})^\alpha}\right)
\end{equation}
for any $\epsilon>0$ and a constant $K_3>0$ w.h.p, where $n^{-\epsilon}$ is the loss in performance due to cooperation overhead. The quantity $n\frac{GP}{N_0W\,(\sqrt{A})^\alpha}$ is the total power received by a node in the receive cluster, when nodes in the transmit cluster are signalling independently at full power. Expressing $T_{HC}$ in terms of $\SNR_s$ in (\ref{SNR_0}), we have
$$
T_{HC}\geq K_3\,n^{1-\epsilon}\,\log\left(1+ n^{1-\alpha/2}\,\SNR_s\right).
$$
Thus, the scaling exponent of hierarchical cooperation is given by
\begin{equation}\label{expo_hier}
e_{HC}(\alpha, \beta)=\left
\{ \begin{array}{ll} 1 & \beta \geq \alpha/2-1 \\
2-\alpha/2+\beta & \beta < \alpha/2-1.
\end{array} \right .
\end{equation}
The performance in the second line is achieved by using a bursty version of the hierarchical cooperation scheme, where nodes operate the original scheme only a fraction $\frac{1}{n^{\alpha/2-1}}$ of the total time and stay inactive in the rest to save power. See~\cite[Sec.~V-A]{OLT07}. 
We see that hierarchical cooperation meets the upper bound in (\ref{upp_expo}) in the first regime when $\beta \geq \alpha/2-1$, i.e., when power is not a limitation. When power is limited but $2\leq \alpha\leq 3$, bursty hierarchical cooperation can be used to achieve the optimal power transfer.
We see that neither multihop nor hierarchical cooperation is able to meet the upper bound in the fourth regime.

\subsection{A New Hybrid Scheme: Cooperate Locally, Multihop Globally}\label{sec:ach:new}

Let us divide our network of $2n$ nodes and area $2\sqrt{A}\times\sqrt{A}$ into square cells of area $A_c=\frac{2A}{2n}\,\SNR_s^{1/(\alpha/2-1)}$. Note that $A_c\leq A$, hence this is a valid choice, if $\beta\leq \alpha/2-1$. If also $\beta>0$, each cell contains of the order of $M=\SNR_s^{1/(\alpha/2-1)}$ nodes w.h.p. We transmit the traffic between the source-destination pairs in the network by multihopping from one cell to the next. More precisely let the S-D line associated to an S-D pair be the line connecting its source node to its destination node. Let the packets of this S-D pair be relayed along adjacent cells on its S-D line just like in standard multihop. See Fig~\ref{fig:mimo_multihop}. The total traffic through each cell is that due to all S-D lines passing through the cell, which is $O(\sqrt{nM})$. Let us randomly associate each of these $O(\sqrt{nM})$ S-D lines passing through a cell with one of the $M$ nodes in the cell, so that each node is associated with $O(\sqrt{n/M})$ S-D lines. The only rule that we need to respect while doing this association is that if an S-D line starts or ends in a certain cell, then the node associated to the S-D line in this cell should naturally be its respective source or destination node. The nodes associated to an S-D line are those that will decode, temporarily store and forward the packets of this S-D pair during the multihop operation. The following lemma states a key result regarding the rate of transmission between neighboring cells.

\begin{lemma}\label{lem:multiHC}
There exists a strategy (based on hierarchical cooperation) that allows each node in the network to relay its packets to their respective destination nodes in the adjacent cells at a rate
$$
R_{relay}\geq K_4\, n^{-\epsilon}
$$
for any $\epsilon>0$ and a constant $K_4>0$.
\end{lemma}

In steady-state operation, the outbound rate of a relay node given in the lemma should be shared between the $O(\sqrt{n/M})$ S-D lines that the relay is responsible for. Hence, the rate per S-D pair is given by
\begin{equation}\label{rate:mhc}
R\geq K_4 \sqrt{M}\, n^{-1/2-\epsilon}
\end{equation}
or equivalently, the aggregate rate achieved by the scheme is
$$
T_{multihop+HC}\geq K_4\, n^{1/2-\epsilon}\, \SNR_s^{\frac{1}{\alpha-2}}.
$$
In terms of the scaling exponent, we have
$$
e_{multihop+HC}(\alpha,\beta)= 1/2+\beta/(\alpha-2) \hspace*{0.3cm}\textrm{if } 0<\beta\leq \alpha/2-1
$$
which matches the upper bound (\ref{upp_expo}) in the third regime.
\medbreak
Note that considering (\ref{rate:mhc}), it is beneficial to choose $M$ as large as possible since it reduces the relaying burden. However, Lemma~\ref{lem:multiHC} does not hold for any arbitrary $M$. The proof of the lemma reveals a key property regarding our initial choice for $M$ (or $A_c$).

{\it Proof of Lemma~\ref{lem:multiHC}:} Let us concentrate only on two neighboring cells in the network. (Consider for example the two cells highlighted in Fig.~\ref{fig:mimo_multihop}): The two neighboring cells together form a network of $2M$ nodes randomly and uniformly distributed on a rectangular area $2\sqrt{A_c}\times\sqrt{A_c}$. Let the $M$ nodes in one of the cells be sources and the $M$ nodes in the other cell be destinations and let these source and destination nodes be paired up randomly to form $M$ S-D pairs. (This traffic will later be used to model the hop between two adjacent cells.) As we have already discussed in (\ref{hc}), using hierarchical cooperation one can achieve an aggregate rate
\begin{align*}
M\, R_{relay}\,&\geq K_3\,M^{1-\epsilon}\,\log\left(1+ M\frac{GP}{N_0W\,(\sqrt{A_c})^\alpha}\right)\\
&=K_3\,M^{1-\epsilon}\,\log\left(1+ M^{1-\alpha/2}\SNR_s\right)
\end{align*}
for these $M$ source destination pairs. The second equation is obtained by substituting $A_c=MA/n$ and $\SNR_s=\frac{GP}{N_0W(A/n)^{\alpha/2}}$. Note that if $M^{1-\alpha/2}\SNR_s\geq 1$, then
\begin{equation}\label{rate:mhc2}
R_{relay}\,\geq\, K_3\,M^{-\epsilon}\,\geq\, K_3\,n^{-\epsilon}.
\end{equation}
In other words, $M=\SNR_s^{1/(\alpha/2-1)}$ is the largest cell size one can choose while still maintaining almost constant transmission rate for each of the $M$ S-D pairs.

Now let us turn back to our original problem concerning the steady-state operation of the multihop scheme. At each hop, each of the $M$ nodes in a cell needs to relay its packets to one of the four (left, right, up and down) adjacent cells. Since the S-D lines are randomly assigned to the nodes in the cell, there are $M/4$ nodes on the average that want to transmit in each direction. These transmissions can be realized successively using hierarchical cooperation and the relaying rate in (\ref{rate:mhc2}) can be achieved in each transmission. On the other hand the TDMA between the four transmissions will reduce the overall relaying rate by a factor of $4$. Indeed, one should also consider a TDMA scheme between the cells allowing only those cells that are sufficiently separated in space to operate simultaneously so that the inter-cell-interference in the network does not degrade the quality of the transmissions significantly. The inter-cell-interference and TDMA will further reduce the rate in (\ref{rate:mhc2}) by a constant factor however will not affect the scaling law. Such insights on scheduling and interference are standard by-now and are not central to our analysis on scaling laws. We refer the reader to \cite[Lemma~4.2]{OLT07} for more details.\hfill$\square$
\medbreak
Note that the new scheme illustrated in Fig.~\ref{fig:mimo_multihop} is a combination of multihop and hierarchical cooperation. Packets are transferred by multihopping on the network level and each hop is realized via distributed MIMO transmissions. Our analysis shows that multihopping and distributed MIMO are two fundamental strategies for wireless networks. However, optimality can only be achieved if these two strategies are combined together appropriately; the optimal combination depends on the SNR level in the network. When $\alpha>3$, we identify three different regimes in wireless networks: the high, low and hybrid SNR regimes. The high SNR regime ($\beta\ge \alpha/2-1$) is the extremal case when even the long-distance SNR in the network is large ($\SNR_l\gg 0$ dB). Distributed MIMO with hierarchical cooperation achieves capacity in this case. In the hybrid SNR regime ($0<\beta\le \alpha/2-1$), the long-distance SNR in the network is low ($\SNR_l\ll 0$ dB), and packets need to be transmitted by multihopping at this scale; while close by pairs are still in the high SNR regime ($\SNR_s\gg 0$ dB) and distributed MIMO provides the optimal information transfer at this smaller scale. The low SNR regime ($\beta\le 0$) is the other extreme when even the short distance SNR is low ($\SNR_s\ll0$ dB). The multihop MIMO scheme reduces to pure multihop in this last case.

\section{Conclusion}\label{sec:conc}
Suppose you are asked to design a communication scheme for a particular network with given size, area, power budget, path loss exponent, etc. What would be the efficient strategy to operate this wireless network? In this paper, we answer this question by connecting engineering quantities that can be directly measured in the network to the design of good communication schemes.  In a given wireless network, we identify two SNR parameters of importance, the short-distance and the long-distance SNR's. The short-distance SNR is the SNR between nearest neighbor pairs. The long-distance SNR is the SNR between farthest nodes times the size of the network. If the long-distance SNR is high, then the network is in the bandwidth limited regime. Long-distance communication is feasible and good communication schemes should exploit this feasibility. If the long-distance SNR is low, then the network is power-limited and good communication schemes need to maximize the power transfer across the network. When the power path loss exponent is small so that signals decay slowly, this power transfer is maximized by global cooperation. When the power path loss exponent is large and signals decay fast, the power transfer is maximized by cooperating in smaller scales. The cooperation scale is dictated by the power path loss exponent and the short-distance SNR in the network.

The current results in the literature, in particular \cite{OLT07} that provides the complete picture for the dense and the extended scaling regimes, fail to answer this engineering question because they only address two specific cases that couple the degrees of freedom and power in the network in two very particular ways. The picture is much richer than what can be delineated by these two settings. In that sense, the current paper suggests the abandonment of the existing formulation of wireless networks in terms of dense and extended scaling regimes, a formulation that has been dominant in the literature over the last decade. A better delineation is obtained by treating the power and degrees of freedom available in the network as two independent parameters and studying the interplay between them.

\appendices

\section{Removing the Assumption of an Empty Strip in Section~\ref{sec:ub}}\label{sec:app2}
While proving the upper bound on network capacity in Section~\ref{sec:ub}, we have considered a vertical cut of the network that divides the network area into two equal halves and assumed that there is an empty rectangular region to the right of this cut, of width equal to the nearest neighbor distance in the network (or of width equal to $1$ in the corresponding rescaled network). With high probability, this assumption does not hold in a random realization of the network. Indeed for any linear cut of the random network, w.h.p. there will be nodes on both sides of the cut that are located at a distance much smaller than the nearest neighbor distance to the cut. In order to prove the result in Section~\ref{sec:ub} rigorously for random networks, we need to consider a cut that is not necessarily linear but satisfies the property of having no nodes located closer than the nearest neighbor distance to it. Below, we show the existence of such a cut using methods from percolation theory. See \cite{FDT06} for a more general discussion of applications of percolation theory to wireless networks.

\begin{figure}[tbp]
\begin{center}
\input{percolation22.pstex_t}
\end{center}
\caption{The cut in Lemma~\ref{lem:percol} that is free of nodes on both sides up to distance $c/2$ is illustrated in the figure.}
 \label{fig:percolation}
\end{figure}
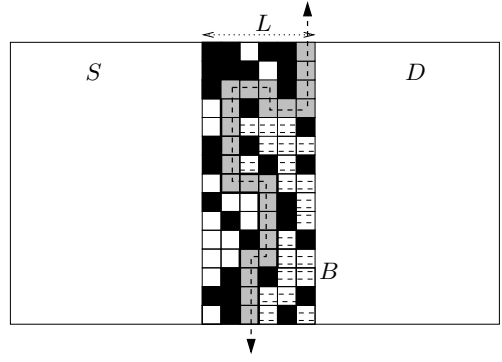

\begin{lemma}
\label{lem:percol}
For any realization of the random network  and a constant $0<c<1/7\sqrt{e}$ independent of $n$ and $A$, w.h.p. there exists a vertical cut of the network area that is not necessarily linear but is located in the middle of the network in a slab not wider than $L=c\sqrt{A/n}\log n$ and is such that there exists no nodes at distance smaller than $\frac{c}{2}\sqrt{A/n}$ to the cut on both sides. See Fig.~\ref{fig:percolation}. 
\end{lemma}

The assumption of an empty region $E$ in Section~\ref{sec:ub}, allowed us to plug in $\hat{w}=1$ in the fourth line of (\ref{SNRtot})
and conclude that when the left-hand side nodes $S$ are transmitting independent signals, the total SNR received by all nodes $D$ to the right of the linear cut is bounded above by 
\begin{align}
\SNR_{tot}&=\sum_{i\in D}\SNR_i\nonumber\\
&\leq \left \{
\begin{array}{ll}
K_1\,\SNR_{s}\, n\, (\log n)^3 & \alpha =2 \vspace*{0.1cm}\\
K_1\,\SNR_{s}\, n^{2-\alpha/2}(\log n)^2  &2 < \alpha <3 \vspace*{0.1cm}\\
K_1\,\SNR_{s}\,\sqrt{n} \, (\log n)^3 & \alpha=3\vspace*{0.1cm} \\
K_1\,\SNR_{s}\,\sqrt{n} \,(\log n)^2 & \alpha > 3, \end{array} \right.\label{SNRtot_lcut}
\end{align}
where $\SNR_i$ is defined in (\ref{SNRi}).

The same result can be proven for the cut given in Lemma~\ref{lem:percol} without requiring any special assumption. Let $B$ denote the set of nodes located to the right of the cut but inside the rectangular slab mentioned in the lemma. See Figure~\ref{fig:percolation}. Then
\begin{equation}\label{eq:sumB}
\SNR_{tot}=\sum_{i\in B}\SNR_i\,+\,\sum_{i\in D\setminus B}\SNR_i.
\end{equation}
For any node $i\in B$, an approximate upper bound for $\SNR_i$ is 
$$
\SNR_i\lesssim\SNR_s\int_{0}^{\sqrt{2\pi}}\int_{c}^{\sqrt{n}}\frac{1}{\hat{r}^\alpha} \hat{r}d\hat{r}d\theta,
$$
since Lemma~\ref{lem:percol} guarantees that there are no left-hand side nodes located at rescaled distance smaller than $c$ to a right-hand side node $i$. Moreover, nodes are uniformly distributed on the network area so the summation in (\ref{SNRi}) over the left-hand side nodes $S$ can be approximated by an integral. A precise upper bound on $\SNR_i$ can be found using the binning argument in \cite[Lemma~5.2]{OLT07} which yields
$$
\SNR_i \leq K_1\,\SNR_s\,\log n.
$$
Since there are less than $\sqrt{n}\log n$ nodes in $B$ with high probability, the first summation in (\ref{eq:sumB}) can be upperbounded by
$$
\sum_{i\in B}\SNR_i\leq K_1\,\SNR_s\,\sqrt{n}\,(\log n)^2.
$$ 
Note that this contribution is smaller than any of the terms in (\ref{SNRtot_lcut}). The second summation $\sum_{i\in D\setminus B}\SNR_i$ in (\ref{eq:sumB}) is equal or smaller in order to (\ref{SNRtot_lcut}) since when the nodes $B$ are removed there is a empty region of width at least $c$ between the nodes $S$ and remaining nodes $D\setminus B$. Hence for the second term in (\ref{eq:sumB}), we are back in the situation discussed in Section~\ref{sec:ub}, hence the upperbound (\ref{SNRtot_lcut}) applies.

{\it Proof of Lemma~\ref{lem:percol}:} Let us divide our network of area $2\sqrt{A}\times\sqrt{A}$ into square cells of side length $c\sqrt{A/n}$ where $0<c<1$ is a constant independent of $A$ and $n$. We say that a cell is closed if it contains at least one node and open if it contains no nodes. Since the $2n$ nodes are uniformly and independently distributed on the network area $2A$, the probability that a given cell is closed is upper bounded by the union bound by
$$
\PP[\textrm{a cell is closed}]\leq c^2.
$$
Similarly, the probability that a given set of $m$ cells $\{c_1,\dots,c_m\}$ are simultaneously closed is upper bounded by
\begin{align}
\PP[&\{c_1,\dots,c_m\}\textrm{ is closed}]\nonumber\\
&=\PP[c_1\textrm{ is closed}]\times\PP[c_2\textrm{ is closed}|c_1\textrm{ is closed}]\times\dots\nonumber\\
&\leq c^2\times c^2\dots\times c^2\,=c^{2m}\label{eq:pn}
\end{align}
since by the union bound we have,
\begin{align*}
\PP[c_{k+1}\textrm{ is closed}&|c_1,\dots, c_{k}\textrm{ is closed}]\\
&\leq \frac{(c^2A/n)}{A-k(c^2A/n)} (n-k)\\
&\leq\, c^2
\end{align*}
when $0<c<1$.
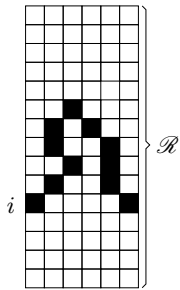
\begin{figure}[tbp]
\begin{center}
\input{percolation3.pstex_t}
\end{center}
\caption{A closed left-right crossing.}
 \label{fig:percolation2}
\end{figure}

Now let us consider a slab of width $c\sqrt{A/n}\log n$ in the middle of the network. Equivalently, this is a rectangle of $\log n\times \sqrt{n}/c$ cells. By choosing $c$ properly, we will show that this slab contains at least one {\em open path} that crosses the network from top to bottom. Such a path is called an open top-bottom crossing. A path is called open if it is composed of neighboring cells that are open, a neighboring cell being one of the four cells located immediately to the top, bottom, left and right of a cell. See Fig.~\ref{fig:percolation}. On the other hand, we define a closed path in a slightly different manner: A closed path is composed of neighboring cells that are closed but a neighboring cell can now be one of the $8$ cells located immediately at the top, top-left, left, bottom-left, bottom, bottom-right, right, top-right of a cell. See Fig.~\ref{fig:percolation2}. With these definitions of closed and open paths, we have
\begin{align*}
\PP[&\textrm{the slab contains an open top-bottom crossing}]\\
&=1-\PP[\textrm{the slab contains a closed left-right crossing}]
\end{align*}
where a closed left-right crossing refers to a closed path that connects the left-boundary $\mathscr{L}$ of the slab to its right boundary $\mathscr{R}$. Let $\PP(i\leftrightarrow\mathscr{R})$ denote the probability that there exists a closed path starting from a particular cell $i\in\mathscr{L}$ and ending at the right-boundary. Note that such a path should be at least of length $\log n$ cells. Denoting by $N_i$ the number of closed paths of length $\log n$ that start from the cell $i$, we have
$$
\PP(i\leftrightarrow\mathscr{R})\leq\PP(N_i\geq 1).
$$
By (\ref{eq:pn}), a given path of length $\log n$ is closed with probability less than $c^{2\log n}$. By the union bound, we have
$$
\PP(N_i\geq 1)\leq c^{2\log n} \sigma_i(\log n),
$$
where $\sigma_i(\log n)$ denotes the number of distinct, loop-free paths of length $\log n$ starting from $i$. This number is obviously not larger than
$
\sigma_i(\log n)\leq 5\times 7^{(\log n-1)}.
$
Combining the three inequalities, we have
\begin{align*}
\PP[&\textrm{the slab contains a closed left-right crossing}]\\
&\leq\sum_{i=1}^{\sqrt{n}/c} \PP(i\leftrightarrow\mathscr{R})
\,\leq\frac{5}{7c}\sqrt{n} (7c^2)^{\log n}. 
\end{align*}
Choosing $c^2<\frac{1}{7\sqrt{e}}$, the last probability decreases to $0$ as $n$ increases. This concludes the proof of the lemma. \hfill$\square$.

\end{document}

%% file: fourregimes.pstex_t
\begin{picture}(0,0)%
\includegraphics{fourregimes.pstex}%
\end{picture}%
\setlength{\unitlength}{2072sp}%
\begingroup\makeatletter\ifx\SetFigFont\undefined%
\gdef\SetFigFont#1#2#3#4#5{%
  \reset@font\fontsize{#1}{#2pt}%
  \fontfamily{#3}\fontseries{#4}\fontshape{#5}%
  \selectfont}%
\fi\endgroup%
\begin{picture}(3765,3126)(1111,-3424)
\put(1846,-2266){\makebox(0,0)[lb]{\smash{{\SetFigFont{7}{8.4}{\familydefault}{\mddefault}{\updefault}{$2$}%
}}}}
\put(2746,-2266){\makebox(0,0)[lb]{\smash{{\SetFigFont{7}{8.4}{\familydefault}{\mddefault}{\updefault}{$3$}%
}}}}
\put(4861,-2221){\makebox(0,0)[lb]{\smash{{\SetFigFont{7}{8.4}{\familydefault}{\mddefault}{\updefault}{$\alpha$}%
}}}}
\put(1981,-2806){\makebox(0,0)[lb]{\smash{{\SetFigFont{7}{8.4}{\familydefault}{\mddefault}{\updefault}{Regime}%
}}}}
\put(1126,-466){\makebox(0,0)[lb]{\smash{{\SetFigFont{7}{8.4}{\familydefault}{\mddefault}{\updefault}{$\beta$}%
}}}}
\put(1171,-2086){\makebox(0,0)[lb]{\smash{{\SetFigFont{7}{8.4}{\familydefault}{\mddefault}{\updefault}{$0$}%
}}}}
\put(4501,-916){\makebox(0,0)[lb]{\smash{{\SetFigFont{7}{8.4}{\familydefault}{\mddefault}{\updefault}{$\beta=\alpha/2-1$}%
}}}}
\put(2296,-961){\makebox(0,0)[lb]{\smash{{\SetFigFont{7}{8.4}{\familydefault}{\mddefault}{\updefault}{Regime I}%
}}}}
\put(2206,-3031){\makebox(0,0)[lb]{\smash{{\SetFigFont{7}{8.4}{\familydefault}{\mddefault}{\updefault}{II}%
}}}}
\put(3286,-1726){\makebox(0,0)[lb]{\smash{{\SetFigFont{7}{8.4}{\familydefault}{\mddefault}{\updefault}{Regime IV}%
}}}}
\put(3286,-2806){\makebox(0,0)[lb]{\smash{{\SetFigFont{7}{8.4}{\familydefault}{\mddefault}{\updefault}{Regime III}%
}}}}
\end{picture}%

%% file: multihopMIMO.pstex_t
\begin{picture}(0,0)%
\includegraphics{multihopMIMO.pstex}%
\end{picture}%
\setlength{\unitlength}{2072sp}%
\begingroup\makeatletter\ifx\SetFigFont\undefined%
\gdef\SetFigFont#1#2#3#4#5{%
  \reset@font\fontsize{#1}{#2pt}%
  \fontfamily{#3}\fontseries{#4}\fontshape{#5}%
  \selectfont}%
\fi\endgroup%
\begin{picture}(5649,3624)(979,-4618)
\put(1261,-4426){\makebox(0,0)[lb]{\smash{{\SetFigFont{7}{8.4}{\rmdefault}{\bfdefault}{\updefault}{$S$}%
}}}}
\put(4366,-2401){\makebox(0,0)[lb]{\smash{{\SetFigFont{7}{8.4}{\rmdefault}{\bfdefault}{\updefault}{$D$}%
}}}}
\end{picture}%

%% file: cutset_small.pstex_t
\begin{picture}(0,0)%
\includegraphics{cutset_small.pstex}%
\end{picture}%
\setlength{\unitlength}{1036sp}%
\begingroup\makeatletter\ifx\SetFigFont\undefined%
\gdef\SetFigFont#1#2#3#4#5{%
  \reset@font\fontsize{#1}{#2pt}%
  \fontfamily{#3}\fontseries{#4}\fontshape{#5}%
  \selectfont}%
\fi\endgroup%
\begin{picture}(9789,8571)(1609,-7924)
\put(8461,-61){\makebox(0,0)[lb]{\smash{{\SetFigFont{6}{7.2}{\rmdefault}{\bfdefault}{\updefault}{$D$}%
}}}}
\put(3601,-61){\makebox(0,0)[lb]{\smash{{\SetFigFont{6}{7.2}{\rmdefault}{\bfdefault}{\updefault}{$S$}%
}}}}
\put(6346,344){\makebox(0,0)[lb]{\smash{{\SetFigFont{6}{7.2}{\familydefault}{\mddefault}{\updefault}{$y$}%
}}}}
\put(7201,-7801){\makebox(0,0)[lb]{\smash{{\SetFigFont{6}{7.2}{\familydefault}{\mddefault}{\updefault}{$\hat{w}$}%
}}}}
\put(11341,-7216){\makebox(0,0)[lb]{\smash{{\SetFigFont{6}{7.2}{\familydefault}{\mddefault}{\updefault}{$x$}%
}}}}
\put(6391,-691){\makebox(0,0)[lb]{\smash{{\SetFigFont{6}{7.2}{\rmdefault}{\bfdefault}{\updefault}{$E$}%
}}}}
\put(7606,-691){\makebox(0,0)[lb]{\smash{{\SetFigFont{6}{7.2}{\rmdefault}{\bfdefault}{\updefault}{$V_D$}%
}}}}
\end{picture}%

%% file: percolation22.pstex_t
\begin{picture}(0,0)%
\includegraphics{percolation22.pstex}%
\end{picture}%
\setlength{\unitlength}{2072sp}%
\begingroup\makeatletter\ifx\SetFigFont\undefined%
\gdef\SetFigFont#1#2#3#4#5{%
  \reset@font\fontsize{#1}{#2pt}%
  \fontfamily{#3}\fontseries{#4}\fontshape{#5}%
  \selectfont}%
\fi\endgroup%
\begin{picture}(5874,4274)(3184,-3998)
\put(6121,-106){\makebox(0,0)[lb]{\smash{{\SetFigFont{9}{10.8}{\familydefault}{\mddefault}{\updefault}{$L$}%
}}}}
\put(4096,-691){\makebox(0,0)[lb]{\smash{{\SetFigFont{9}{10.8}{\rmdefault}{\bfdefault}{\updefault}{$S$}%
}}}}
\put(7921,-691){\makebox(0,0)[lb]{\smash{{\SetFigFont{9}{10.8}{\rmdefault}{\bfdefault}{\updefault}{$D$}%
}}}}
\put(6886,-3076){\makebox(0,0)[lb]{\smash{{\SetFigFont{9}{10.8}{\rmdefault}{\bfdefault}{\updefault}{$B$}%
}}}}
\end{picture}%

%% file: percolation3.pstex_t
\begin{picture}(0,0)%
\includegraphics{percolation3.pstex}%
\end{picture}%
\setlength{\unitlength}{2072sp}%
\begingroup\makeatletter\ifx\SetFigFont\undefined%
\gdef\SetFigFont#1#2#3#4#5{%
  \reset@font\fontsize{#1}{#2pt}%
  \fontfamily{#3}\fontseries{#4}\fontshape{#5}%
  \selectfont}%
\fi\endgroup%
\begin{picture}(1830,3399)(5251,-3628)
\put(7066,-1996){\makebox(0,0)[lb]{\smash{{\SetFigFont{9}{10.8}{\familydefault}{\mddefault}{\updefault}{$\mathscr{R}$}%
}}}}
\put(5266,-2716){\makebox(0,0)[lb]{\smash{{\SetFigFont{9}{10.8}{\familydefault}{\mddefault}{\updefault}{$i$}%
}}}}
\end{picture}%